# Unveiling the impact of anti-site defects in magnetic transitions of few-layer MnBi$_2$Te$_4$ by operando heating

Xinyu Chen[1†], Jingjing Gao[1†*], Shuang Wu[1†*], Zhiwei Huang[1†], Zhongxun Guo[1], Canyu Hong[1], Ruohan Chen[1], Mingyan Luo[1], Zhaochen Liu[1], Zeyuan Sun[1,2,3], Wei Ruan[1,4], Jing Wang[1,2,5], Yuanbo Zhang[1,2,4,5,6*], Shiwei Wu[1,2,3,5*]

[1] State Key Laboratory of Surface Physics and Department of Physics, Fudan University, Shanghai 200433, China

[2] Institute for Nanoelectronic Devices and Quantum Computing, Fudan University, Shanghai 200433, China

[3] Key Laboratory of Micro and Nano Photonic Structures (MOE), Fudan University, Shanghai 200433, China

[4] Shanghai Branch, Hefei National Laboratory, Shanghai 201315, China

[5] Shanghai Research Center for Quantum Sciences, Shanghai 201315, China

[6] New Cornerstone Science Laboratory, Shenzhen 518054, China

[†] These authors equally contributed to this work.

[*] Corresponding emails: gaojj@fudan.edu.cn, swu17@fudan.edu.cn, zhyb@fudan.edu.cn, swwu@fudan.edu.cn

**Abstract**

As the first experimentally discovered intrinsic magnetic topological insulator, MnBi$_2$Te$_4$ has attracted widespread attentions, providing a unique platform for the exploration of topological quantum phases, such as quantum anomalous Hall effect and axion insulator state. Despite the increasing number of potential factors affecting samples being identified, obtaining the high-quality device performance with desired topological quantum phases remains a challenge. In this work, by comparing the reflective magnetic circular dichroism (RMCD) of crystals with different defect densities that are characterized by atomically resolved scanning tunneling microscopy, we demonstrate that anti-site defects play an essential role in achieving ideal magnetic states. By measuring RMCD hysteresis loops with operando heating, we find that MnBi$_2$Te$_4$ few-layer samples are highly susceptible to thermal impact, even at temperature as low as 45°C. The magnetic behavior of heating-treated samples is akin to that of samples fabricated into devices, revealing the thermal impact on devices as well. Starting from few-layers with ideal layer-dependent magnetic order, thermal heating leads to the convergence of magnetization and transition fields between odd- and even-layers. The observed heating-induced magnetic evolution can serve as a valuable reference for assessing the sample quality or the density of anti-site defects. Our findings not only point out the long-standing hidden factor that arose controversies in MnBi$_2$Te$_4$, but also pave the way for controllably engineering the topological quantum phenomena.

**Introduction**

The interplay between magnetism and topology allows for the exploration of exotic quantum phenomena including the quantum anomalous Hall (QAH) effect[1-7] and the axion insulator state[8-13], which have been investigated in magnetically doped topological insulator and intrinsic magnetic topological insulator. $MnBi_2Te_4$, as the first intrinsic magnetic topological insulator with better magnetic homogeneity, sparks extensive research interest[5-7,11-24]. As shown in Fig. 1a, its crystal structure consists of septuple layers (SLs). Figure 1b depicts its A-type antiferromagnetic structure, giving rise to layer-dependent topological states: Odd-number-SL $MnBi_2Te_4$ with uncompensated magnetization leads to the QAH effect, while even-number-SL $MnBi_2Te_4$ with zero net magnetization is predicted to achieve the axion insulator phase[16-18].

Despite tremendous progress in recent years, further exploration of the topological quantum states in $MnBi_2Te_4$ remains a challenge. For odd-number-SL $MnBi_2Te_4$, the anomalous Hall effect often exhibits varying degrees of quantized Hall conductance across different samples[6,20,24-27]. The deviation from quantization is clearly exemplified on a 5SL $MnBi_2Te_4$ device (Fig. 1c). For even-number-SL $MnBi_2Te_4$, odd-layer-like uncompensated magnetic hysteresis loops are often observed both in magneto-optical spectroscopy[26,28] and transport[5,26,27,29] in previous studies, hindering the realization of axion insulator state. We also observed the unexpected loop in the transport measurement of a 4SL device in Fig. 1d. Various external factors are identified to hamper the emergence of ideal topological states, such as surface degradation by air exposure[30] and polymer coatings[27]. However, even maintaining fabrication process isolated from air and chemicals, these devices still exhibit nonideal states both in odd- and even-layers (Fig. 1c and d). This dilemma requires further identification of crucial factors that suppress the observation of ideal topological states.

In this work, by comparing RMCD hysteresis loops from crystals with varying defect densities that are characterized by atomically resolved scanning tunneling microscopy, we unambiguously identify density of anti-site defects as a crucial factor in determining magnetic ground state and therefore the topological state. Beginning with ideal layer-dependent magnetism in MnBi$_2$Te$_4$ few-layers with low defect density, upon operando heating, the net magnetization and transition fields of the odd- and even-layers progressively converge, diminishing the contrasting odd-even-layer dependence. As the magnitude of spin-flip transition fields observed in heating-treated samples reproduces the magnetic behaviors found in fabricated devices, we point out that the nonideal behaviors in devices result from the effective heating during the electrode evaporation process, which is a commonly-used but long-overlooked factor. Our results reveal the thermal impact on the behaviors of magnetic topological states, shedding light on the manipulation of magnetism-topology interactions in this intriguing material.

**Results**

All MnBi$_2$Te$_4$ few-layers used in this work were mechanically exfoliated onto SiO$_2$/Si substrate using adhesive tape (Methods). A typical microscopic image of MnBi$_2$Te$_4$ thin flake with different thickness is shown in Supplementary Fig. 1a. The number of layers was precisely determined by their optical contrast combined with atomic force microscopy[31] (Supplementary Fig. 1b). The sample preparation process was conducted in an argon-filled glove box, including the device fabrication for MnBi$_2$Te$_4$ samples with electrodes described later in this work. The samples were sealed and transferred to our magneto-optical cryostat in high vacuum for the RMCD and transport measurements.

We performed RMCD measurements on MnBi$_2$Te$_4$ few-layers exfoliated from two types of crystals. Type-A crystals were grown by an optimized flux method, whereas Type-B crystals were grown under normal flux method (Methods). Figure 2a and 2b show the RMCD hysteresis of freshly exfoliated flakes obtained from type-A and type-B crystals, with the layer numbers ranging from 4SL to 7SL. For the spin-flop transition between the collinear AFM and canted AFM states, samples from both crystal types exhibit the even-odd-layer oscillation of transition fields. As labeled by $H_{c2}$ with green arrows in Fig. 2a and 2b, the transition field is about 2 T for even-layer flakes and 3.5 T for odd-layer flakes. For the spin-flip transition between the collinear AFM states (labeled by $H_{c1}$ with orange arrows in Fig. 2a and 2b), however, flakes from different crystal types exhibit distinct features: even-layer samples from Type-B crystals show hysteresis loops, whereas samples from Type-A crystals do not. In samples from Type-A crystals, as the magnetic field sweeping upward or downward, the odd-layer $MnBi_2Te_4$ samples exhibit a sudden jump, while no spin-flip transition can be identified in the even-layer ones. This finding is consistent with the theoretically proposed A-type antiferromagnetic structure[16-18]. In stark contrast to even-odd-layer dependent magnetic states, for flakes from Type-B crystals, both even- and odd-layers display prominent hysteresis loops, which is similar to previously reported optical[26,28] or electrical transport results[5,26,27,29].

To uncover the underlying mechanisms behind the distinct magnetic ground states observed in different types of crystals, we employed scanning tunneling microscopy (STM) for detailed characterizations. As shown in Fig. 2c and 2d, both samples exhibit two types of defects, including the dark triangular $Mn_{Bi}$ anti-site defect and the bright circular $Bi_{Te}$ anti-site defect[32-35] (see Supplementary Fig. 2 for detailed defect structures), but the defect density is much lower in type-A crystals compared to type-B. In the type-A crystals grown by the optimized synthesis, the concentration of $Mn_{Bi}$ anti-site defects is about 1.9%, which is almost four times sparser than that in crystals grown by the normal synthesis (7%). Moreover, the density of $Bi_{Te}$ defects is 0.15% and 0.79% in the optimized and normal flux method grown crystals, respectively. The correlation between defect density and hysteretic response in $MnBi_2Te_4$ few-layers suggests that the magnetic order is modified by the defects. As previously proposed[33,36,37], Mn atoms in the Bi layer can introduce local magnetic moment with its orientation antiparallel to that of the Mn layer, leading to the uncompensated hysteresis loop. Therefore, controlling the defect density during growth is crucial for high-quality crystals with ideal magnetic structure[38-40].

Utilizing optimized crystals with low defect density and ideal magnetic states, we fabricated $MnBi_2Te_4$ devices and performed simultaneous transport and RMCD measurements to investigate the magnetic topological states. Figure 3a and 3b show the longitudinal ($R_{xx}$) and transverse ($R_{yx}$) resistances as a function of magnetic field in a typical 5SL $MnBi_2Te_4$ device at various back-gate voltages ($V_g$). The sample was measured at 6 K, well below its Néel temperature of about 23 K (Supplementary Fig. 3a). The quantized Hall resistance $R_{yx} = \pm h/e^2$ and nearly vanishing longitudinal resistance $R_{xx}$ in high magnetic field regions denote a Chern insulator state with Chern number ($C$) of ±1 at $V_g$ = -30 V, where the Fermi level is close to the charge neutral point (CNP) within the surface gap (Supplementary Fig. 3b-d). The reversal of anomalous Hall effect in Fig. 3b may come from the competition between intrinsic Berry curvature and Dirac-gap enhanced extrinsic skew scattering in this material, because the skew scattering contributes a positive anomalous Hall effect and the intrinsic contribution from Berry curvature is negative[41]. In such electrical transport measurements, both spin-flip and spin-flop transitions are captured by the jumps at about ±0.5 T ($Hc_1$) and ±3 T ($Hc_2$), respectively. These transitions also occur in the simultaneously measured magnetic-field dependent RMCD at the exactly same magnetic fields (Fig. 3c). Note that, although the transport data vary with the Femi level tuned by gate voltage, the RMCD results remain independent of the gate voltage. This observation demonstrates that the RMCD retrieves exclusive magnetization information, regardless of the specific electronic properties of materials.

The RMCD also enables the examination of possible domain structures through microscopic imaging that could not be obtained by transport measurements. Figure 3d and 3e-f show the optical microscopy of the 5SL $MnBi_2Te_4$ device and the corresponding RMCD images at zero field, respectively. When the magnetic field was swept upward from -6.5 T (Fig. 3e) or downward from 6.5 T (Fig. 3f), both the RMCD images exhibit a homogeneous intensity across the device, but with the opposite magnetization. The RMCD microscopy suggests that, unlike what was reported for bulk $MnBi_2Te_4$[42,43], the odd-layer device contains a single magnetic domain with non-zero switchable magnetization.

The RMCD and transport measurements were also conducted in a 4SL $MnBi_2Te_4$ device (Fig. 3g-i). Similar to the odd-layer devices, a high-field Chern insulator state is also achieved at gate voltage ranging from -25 V to 15 V, centered around the CNP of -5 V (Supplementary Fig. 4). Strikingly, a spin-flip transition arises in the magnetic-field dependent RMCD loop, distinct from that in the freshly exfoliated 4SL sample presented in Fig. 2a. Such a transition can also be distinguished in the magnetic-field dependent $R_{xx}$ and $R_{yx}$, despite the large noise in $R_{yx}$ at CNP due to the giant longitudinal resistance. The corresponding optical microscopy and RMCD images of the device are shown in Fig. 3j-l, again demonstrating that the emergent magnetic state uniformly exists in the entire device.

By further comparing the details of magnetic transitions between the freshly exfoliated 4SL sample (Fig. 2a) and the fabricated 4SL device (Fig. 3l), two key distinctions can be observed. The most evident one, as mentioned above, is the appearance of spin-flip transitions in the device, suggesting the formation of an additional magnetic order. Moreover, the transition field of the spin-flop transition has expanded from about 1.7 T to 2.0 T. These distinctions suggest a variation in the magnetic ground state of $MnBi_2Te_4$ few-layers after being fabricated into devices. Because we directly deposit electrodes onto samples covered by a stencil mask[5,29], the samples remain free from any polymers or chemical solutions. Such a variation likely originates from the electrode deposition process, which prompts us to investigate the key factor that drives this modification.

To investigate this factor, we performed a series of RMCD measurements at 6 K after the as-exfoliated $MnBi_2Te_4$ few-layers were consecutively heated up to 90 °C inside the magneto-optical cryostat (see Methods for details of the thermal treatment). Figure 4a-d show the RMCD results before and after the consecutive heating processes for 3-6SL, respectively. Variations in both spin-flip and spin-flop transitions can be identified in the odd-layers upon heating. A spin-flip transition emerges in the 4SL and 6SL samples after heating up to 60 °C and 75 °C, respectively. The transition field of their spin-flop transition also changes with heating temperature. These observations suggest that the thermal treatment alters the magnetic structure of $MnBi_2Te_4$ few-layers.

The characteristics of RMCD hysteresis loops in $MnBi_2Te_4$ few-layers are extracted and plotted in Fig. 4e-g. For a direct and fair comparison, all the measured samples are on the same $SiO_2$/Si substrate, and can contain multiple regions with the same thickness. Figure 4e and 4f show the transition fields of the spin-flip ($H_{c1}$) and spin-flop ($H_{c2}$) transitions, respectively, as functions of heating temperature. Here we only extract the positive critical fields, because of the absence of a noticeable exchange bias effect under a symmetric, high magnetic sweep and thus the negligible difference between two sides. For the odd-layers, $H_{c1}$ increases with heating temperature, whereas $H_{c2}$ decreases. By contrast, the even-layers show the opposite trends in both $H_{c1}$ and $H_{c2}$. The same contrast is observed in the remnant RMCD intensity at zero magnetic field (Fig. 4g), which increases monotonically with heating temperature in the even-layers, while it decreases in the odd-layers. Eventually around the heating temperature of 90 °C or higher, the hysteresis loops become alike between the odd- and even-layers, making the identification of layer parity more difficult.

Such magnetic evolution upon heating is further reproduced in multiple samples located on three $SiO_2$/Si substrates (Supplementary Fig. 5). For the transition fields $H_{c1}$ and $H_{c2}$, minor fluctuations among different samples are observed, and can be attributed to the inhomogeneous defects[44,45] or undesirable strains[46,47] in the samples, which result in the field-pinning of domain wall and consequently modify the magnetic transition fields. Meanwhile, the remnant RMCD (Fig. 4g), which is proportional to the out-of-plane magnetization, has better consistency between different samples with the same thickness. Nevertheless, the RMCD results demonstrate the vulnerability of the $MnBi_2Te_4$ few-layers subject to heating, with both the magnetization and transition fields exhibiting significant changes at a heating temperature as low as 45 °C.

Magnetic behavior of the $MnBi_2Te_4$ devices (Fig. 3) is therefore reproduced by heating the as-exfoliated few-layers (Fig. 4), revealing the thermal impact of electrode deposition for device fabrication as the key factor. Moreover, the transition fields ($H_{c1}$ and $H_{c2}$), along with the remnant RMCD intensity, as a function of heating temperature can serve as a reference guide to evaluate the thermal impact of electrode deposition in $MnBi_2Te_4$ devices. Examining the hysteresis loops for the devices shown in Fig. 3, the equivalent heating temperature is estimated to about 90 °C for Au electrodes. When the even-layer 4SL device was made with additional Ti deposition as a wetting layer for Au electrodes, as shown in Supplementary Fig. 6, the hysteresis loops seem more like those of an odd-layer. The equivalent heating temperature thus becomes even higher than 90 °C. This temperature increase is understandable because the Ti deposition imposes more severe thermal impact to the samples than the Au evaporation.

Because the magnetic property of MnBi$_2$Te$_4$ few-layers, particularly the appearance of spin-flip transitions in the even-layers, is directly correlated with the higher density of anti-site defects, as illustrated in Fig. 2, the heating effect might be closely related to the accumulation of such defects. As the top surface layer of MnBi$_2$Te$_4$ is more susceptible to thermal impact, here we note that the defects are likely unevenly distributed between different layers, with the top surface having relatively higher defect density. Our experimental findings would inspire the future studies on how the defect distribution, or more generally the defect engineering, actively affects the exotic quantum phenomena and performance in MnBi$_2$Te$_4$[38-40].

**Conclusions**

In summary, we find that MnBi$_2$Te$_4$ crystals with different defect densities exhibit distinct magnetic transition behaviors under an out-of-plane magnetic field, highlighting the crucial role of anti-site defects in shaping their magnetic properties. By simultaneously conducting transport and RMCD measurements, we establish the connection between electrical transport properties and magnetism in MnBi$_2$Te$_4$ few-layers that conforms to the layer parity. Utilizing RMCD measurements with operando heating, we find the magnetic property of MnBi$_2$Te$_4$ is highly sensitive to heating, even at temperature as low as 45°C. Moreover, the remnant magnetization can serve as a reference for evaluating the sample quality or the density of anti-site defects. These insights are crucial for advancing the development of MnBi$_2$Te$_4$-based quantum devices, and gaining a deeper understanding of its magnetic and topological properties.

**Methods**

**MnBi$_2$Te$_4$ crystal growth and selection**

MnBi$_2$Te$_4$ single crystals were synthesized through the self-flux growth method. High-purity Mn (99.95%), Bi (99.999%), and Te (99.999%) powders with a molar ratio of 1:2:4 were thoroughly mixed and loaded into a silica crucible. The crucible was subsequently sealed within a silica ampoule under vacuum conditions ($<10^{-4}\ mbar$). In the case of type-A crystals, we implemented an optimized growth protocol involving extended isothermal holding at the crystallization temperature combined with a slow annealing process. This modified approach significantly reduced the defect density in the crystals, enabling the production of high-quality MnBi$_2$Te$_4$ samples. Detailed growth parameters and optimization procedures are described in ref. [48]. For the type-B crystals, the mixture was initially heated to 900°C and maintained at this temperature for 24 hours to ensure complete homogenization. It was then subjected to a controlled cooling process, gradually decreasing the temperature to 595°C over 150 hours, followed by immediate centrifugation at this temperature to separate the crystals from the flux.

**Sample preparation and device fabrication**

MnBi$_2$Te$_4$ few-layer samples were obtained by standard mechanical exfoliation onto plasma cleaned Si substrates with 285-nm-thick SiO$_2$ coating. The flake thickness was determined by optical contrast and atomic force microscopy (AFM). Proceeding to the device fabrication, a Hall bar configuration was achieved by initially placing a pre-patterned SiN membrane window over the selected thin flake, serving as a precise stencil mask. We then utilized an electron-beam evaporator to deposit electrode films (60-nm Au or 5-nm Ti/60-nm Au) through the stencil mask. The whole process was carried out in a glove box with argon/nitrogen atmosphere (H$_2$O < 0.1 ppm, O$_2$ < 0.5 ppm). To prevent from air exposure, all the samples were sealed with a cover glass by vacuum grease in custom-built sample holders and transferred from glove box to the magneto-optical cryostat swiftly.

**RMCD and transport measurements with *in situ* heating**

All the RMCD and transport measurements were conducted in a home-built variable-temperature magneto-optical cryostat down to 6 K and up to 7 T. The thermal treatment was performed *in situ* inside the same cryostat, involving: heating the substrate to specific temperatures, maintaining at each temperature for a duration of 2 hours, and cooling down to the base low temperature for RMCD and transport measurements. Most of the measurements were obtained at the base temperature of 6 K, unless otherwise noted. Each sample was sequentially heated to 45°C, 60°C, 75°C, and 90°C because of the sample irreversibility. The entire process was meticulously conducted under high vacuum conditions, with a pressure maintained below $1 \times 10^{-8}$ Torr inside the magneto-optical cryostat.

The RMCD measurements were in the backscattering configuration. The magnetic field was applied perpendicular to the sample plane. A linearly polarized 633 nm HeNe laser (Thorlabs), was used. After passing through an optical chopper (Thorlabs) at 197 Hz, its polarization was modulated by a photoelastic modulator (PEM) at 50.052 kHz. Using a 50× objective (Nikon, NA 0.55), the laser beam was focused onto the sample with the spot diameter of ~2 μm. The excitation power was 1 μW. The reflected light was detected by an Avalanche Photodiode (Thorlabs), and analyzed by a lock-in amplifier (Zurich Instruments). The magnitude of RMCD was determined by the ratio of first-harmonic components at 50.052 kHz and 197 Hz.

Transport measurements were performed by using a standard lock-in technique with an excitation current of 10 ~ 80 nA (Signal recovery 7265). The bottom gate voltage $V_g$ was controlled by a source meter (Keithley 2400). The $R_{xx}$ ($R_{yx}$) data shown in this paper were symmetrized (anti-symmetrized).

**Data availability**

The data supporting the findings of this study are available within the paper. Source data are provided with this paper.

**Acknowledgements**

The work at Fudan University was supported by National Key Research and Development Program of China (Grant Nos. 2022YFA1403300, 2024YFA1409800), National Natural Science Foundation of China (Grant Nos. 12034003, 12427807, 12204115, 1235000130, 12350404), Science and Technology Commission of Shanghai Municipality (Grant Nos. 20JC1415900, 23JC1400400, 23JC1400600, 24LZ1400100, 2019SHZDZX01), Shanghai Municipal Education Commission (Grant No. 2021KJKC-03-61), and China Postdoctoral Science Foundation (Grant No. 2022M720812).

**Author contributions**

S.W.W. and Y.Z. conceived and supervised the project. X.Y.C. and S.W. conducted the transport and RMCD experiments with the assistance of J.J.G., Z.X.G., C.Y.H. and R.H.C. J.J.G., Z.X.G. and M.Y.L. grew the MnBi$_2$Te$_4$ crystals and prepared the few-layer samples/devices. Z.W.H. conducted the STM measurements. X.Y.C., J.J.G., S.W., Y.B.Z. and S.W.W. analyzed the data and wrote the paper with contributions from all authors.

**Competing interests**

The authors declare no competing interests.

**Figures and Captions**

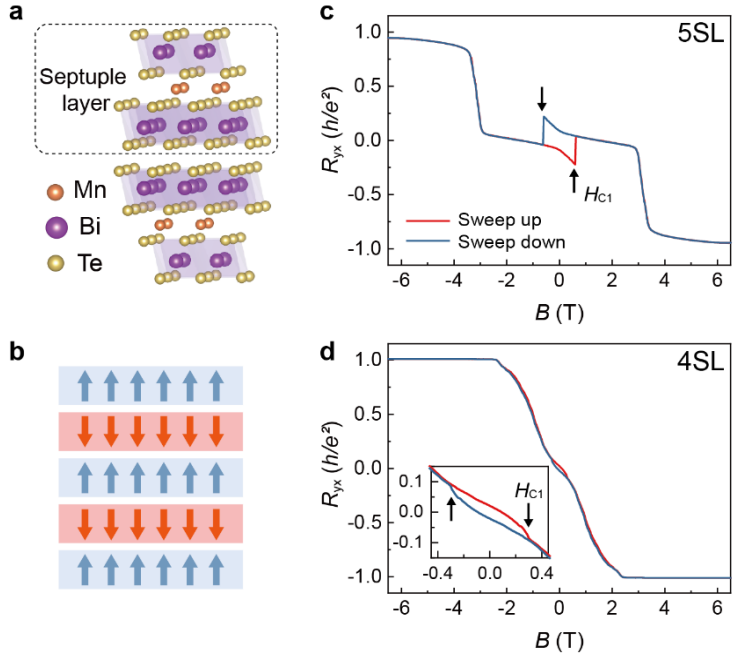

**Figure 1 | Appearance of spin-flip transitions in both odd- and even-layer MnBi$_2$Te$_4$. a**, Crystallographic structure of MnBi$_2$Te$_4$. **b**, Schematic of the theoretically proposed layered antiferromagnetic ground state of MnBi$_2$T$_4$. The spins of the Mn$^{2+}$ ions ferromagnetically couple to each other within one septuple layer, while the adjacent layers are antiferromagnetically coupled. **c-d**, Variable magnetic field transverse resistance ($R_{yx}$) of 5SL and 4SL MBT samples at the temperature of 6 K and 2 K, respectively. The back-gate voltage was tuned near the charge neutral point of the sample. Inset figure in **d** is the zoomed-in section that highlights the hysteresis loop.

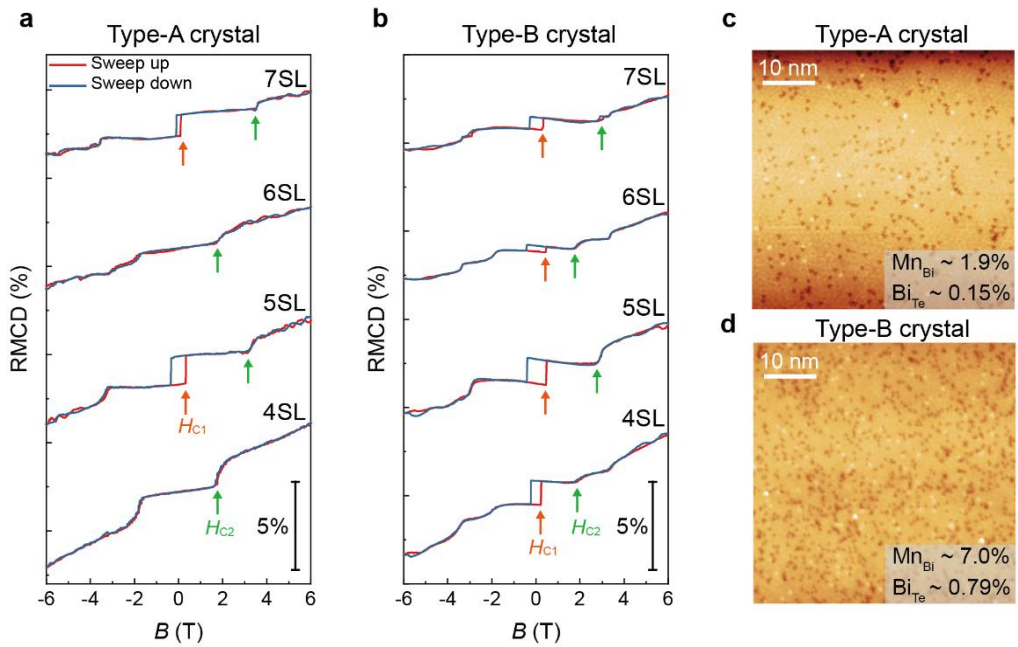

**Figure 2 | Distinct RMCD results in MnBi$_2$Te$_4$ with different defect density. a, b,** RMCD loops of the freshly exfoliated 4SL to 7SL MnBi$_2$Te$_4$ samples exfoliated from the crystals grown by optimized flux method (**a**) and normal flux method (**b**). The measuring temperature is 6 K. Curves are vertically offset for comparison. **c, d,** corresponding STM topographies of the MnBi$_2$Te$_4$ samples exfoliated from crystals grown by optimized flux method (**c**) and normal flux method (**d**), respectively. The concentration of the Mn$_{Bi}$ anti-site defect is about 7% in samples grown by normal flux, which is almost four times larger than that in samples grown by the optimized flux (1.9%). The density of Bi$_{Te}$ defect is 0.15% and 0.79% in the optimized and normal flux method grown samples, respectively. Bias: 1 V, setpoint: 10 pA (**c**) and 100 pA (**d**). The scale bar is 10 nm.

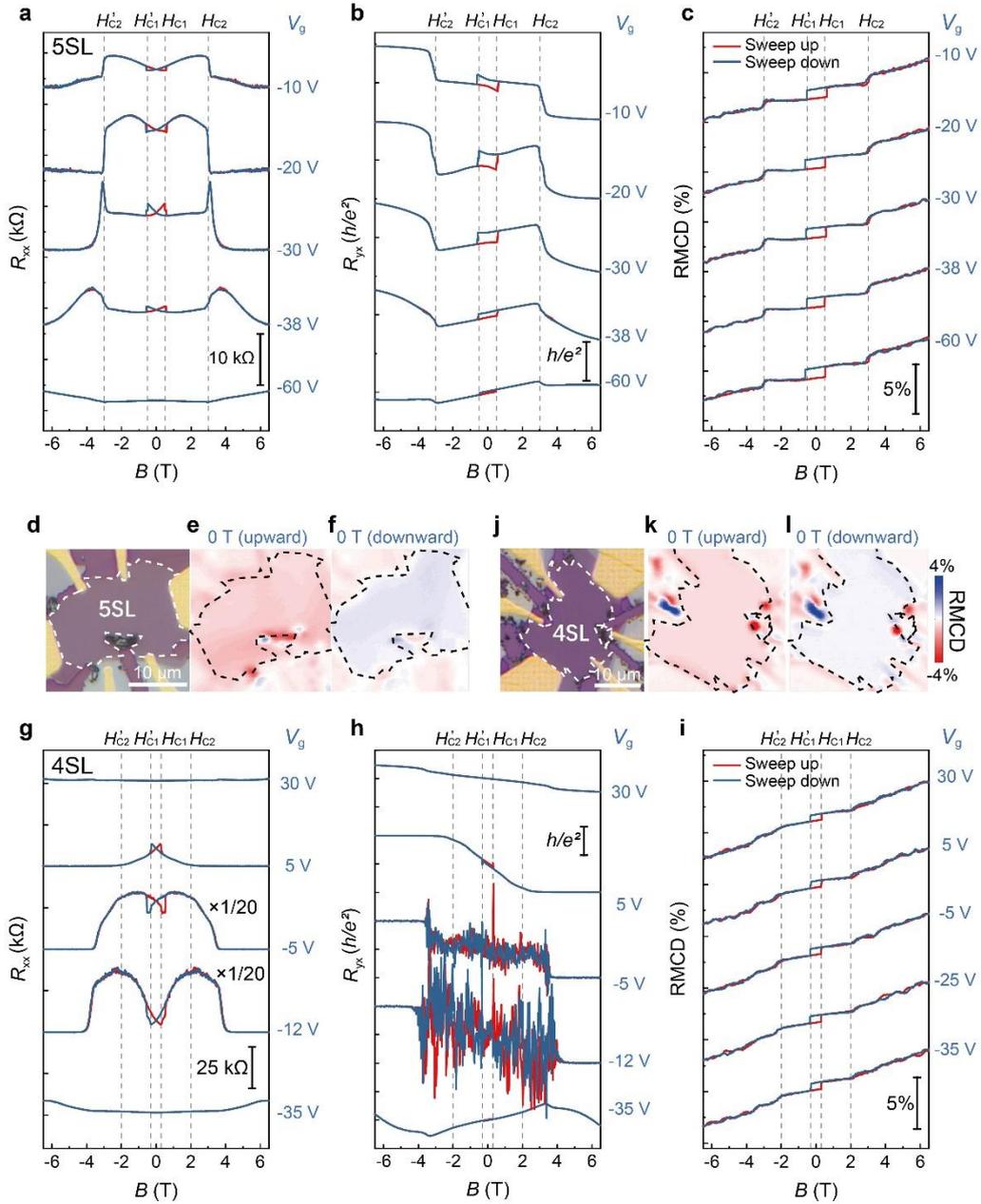

**Figure 3 | Comparison between electrical transport and the corresponding RMCD results in 5SL and 4SL MnBi$_2$Te$_4$ devices. a-c**, Magnetic-field dependence of $R_{xx}$ (**a**), $R_{yx}$ (**b**), and RMCD signal (**c**) for a 5SL device at varying gate voltages. **d**, Optical microscopic image of the 5SL device measured in **a-c**. Metal (Au) contacts to the sample were e-beam deposited through a stencil mask in a Hall bar geometry. **e, f**, RMCD images of the 5SL device taken at zero magnetic field, sweeping from -6.5 T to 0 T (**e**) or from 6.5 T to 0 T (**f**). Both of the images show a single domain structure throughout the sample. **g-i**, Magnetic-field dependence of $R_{xx}$ (**g**), $R_{yx}$ (**h**), and RMCD signal (**i**) for the 4SL device at varying gate voltages. **j-l**, Optical microscopy (**j**) and RMCD images at 0 T (**k**, from -6.5 T and **l**, from 6.5 T) of the 4SL device measured in **g-i**. The distinct magnetizations at zero magnetic field between upward and downward sweeping are captured by the RMCD images.

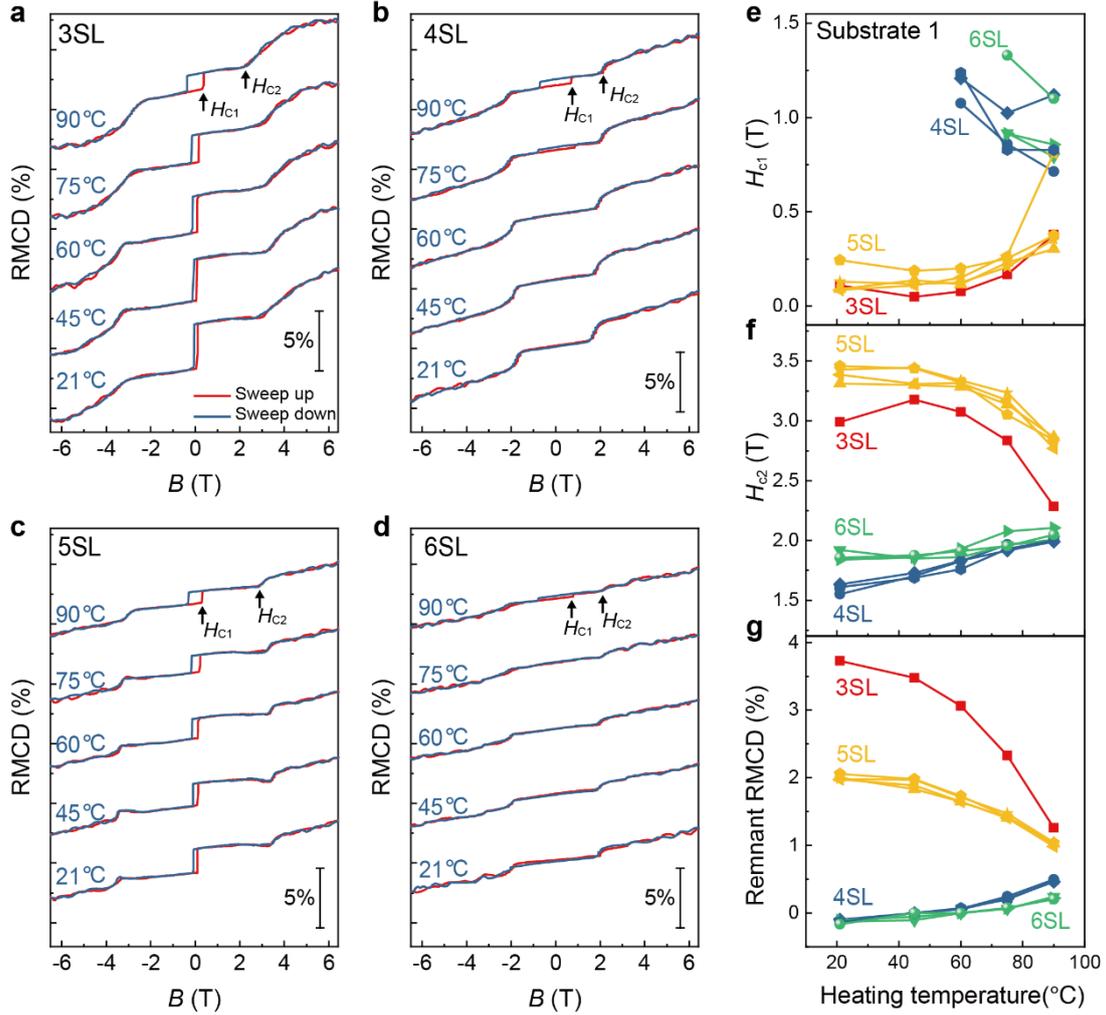

**Figure 4 | Magnetic evolution upon thermal treatment in MnBi$_2$Te$_4$ few-layers. a-d,** Magnetic-field dependent RMCD signals acquired in 3-6 SL MnBi$_2$Te$_4$ before and after heating with temperatures ranging from 45 °C to 90 °C. The samples were exfoliated at room temperature and marked as 21 °C. All of the RMCD data were taken at the temperature of 6 K. **e, f,** The extracted spin-flip ($H_{c1}$, **e**) and spin-flop ($H_{c2}$, **f**) transition field as functions of heating temperature. The transition fields are extracted from the positive-field side of the data, as the difference between positive and negative transition fields is less than 0.1 T. **g,** Remnant RMCD intensity at zero magnetic field as a function of heating temperature. All of the data shown in **e-g** are taken from multiple few-layers located in the same substrate. Several samples with the same number of layers are measured, and denoted by various of symbols. This allows for a direct comparison of the magnetic-field dependent RMCD between samples with the same or different numbers of layers.